\documentclass[conference]{IEEEtran}
\ifCLASSINFOpdf
   \usepackage[pdftex]{graphicx}
\else
\fi
\usepackage{array}

\usepackage{siunitx}
\usepackage{graphicx}
\usepackage{color}
\usepackage{booktabs}
\usepackage{multirow}
\usepackage[table,xcdraw]{xcolor}
\usepackage[font=small,skip=2pt]{caption}

\begin{document}
%
\title{Performance evaluation over HW/SW co-design SoC memory transfers for a CNN accelerator}

\author{\IEEEauthorblockN{A. Rios-Navarro, R. Tapiador-Morales, A. Jimenez-Fernandez, M. Dominguez-Morales, C. Amaya and \\A. Linares-Barranco}
\IEEEauthorblockA{Escuela T\'ecnica Superior de Ingenier\'ia Inform\'atica. 
University of Seville, Seville, Spain\\
arios@atc.us.es    http://www.rtc.us.es/}
}


%


\maketitle

\begin{abstract}
Many FPGAs vendors have recently included embedded processors in their devices, like Xilinx with ARM-Cortex A cores, together with programmable logic cells. These devices are known as Programmable System on Chip (PSoC). Their ARM cores (embedded in the processing system or PS) communicates with the programmable logic cells (PL) using ARM-standard AXI buses. In this paper we analyses the performance of exhaustive data transfers between PS and PL for a Xilinx Zynq FPGA in a co-design real scenario for Convolutional Neural Networks (CNN) accelerator, which processes, in dedicated hardware, a stream of visual information from a neuromorphic visual sensor for classification. In the PS side, a Linux operating system is running, which recollects visual events from the neuromorphic sensor into a normalized frame, and then it transfers these frames to the accelerator of multi-layered CNNs, and read results, using an AXI-DMA bus in a per-layer way. As these kind of accelerators try to process information as quick as possible, data bandwidth becomes critical and maintaining a good balanced data throughput rate requires some considerations. We present and evaluate several data partitioning techniques to improve the balance between RX and TX transfer and two different ways of transfers management: through a polling routine at the user-level of the OS, and through a dedicated interrupt-based kernel-level driver. We demonstrate that for longer enough packets, the kernel-level driver solution gets better timing in computing a CNN classification example. Main advantage of using kernel-level driver is to have safer solutions and to have tasks scheduling in the OS to manage other important processes for our application, like frames collection from sensors and their normalization.

\end{abstract}


%
\IEEEpeerreviewmaketitle

\section{Introduction}

In the field of the embedded systems, SoC chips have played an important role in the evolution of this technology area. The most recent SoC chips have several peripherals that increase the range of applications that can be used for. Some of these peripherals are used for digital, analog, mixed-signal, often radio-frequency functions and recently SoC chips include a graphic dedicated hardware in order to accelerate graphical applications. In recent years, the dominance of SoC for embedded system application begins to be questioned. FPGA (Field Programmable Gate Array) devices with on-chip processing system, known on the literature as SoC FPGA or PSoC (Programmable System on Chip), have recently emerged as potential solutions for compact processing applications. PSoCs combine the better of two world, they have a familiar processing system development interface for sequential algorithms or embedded OS applications, and at the same time, they provide an empty landscape for custom hardware development that enlarges the set of application of our system. PSoCs also offer a flexible programmable alternative for sequential processing, implementing any hardware function to augment the capabilities the PS owns. In fact, due to the inherently parallel nature of the FPGA, multiple hardware blocks can operate simultaneously, either in parallel, when the logic is replicated, or in a pipelined stages. These capabilities open up a wide range of possibilities for applications that can be deployed in these systems. PSoCs can be found in different applications where main and lighter tasks are processed by the PS, while harder computational tasks are designed to be deployed on the PL. Some examples are: 

\textbf{- Automotive}: Cars nowadays contain Advanced Driver Assistance Systems (ADAS) that refers specifically to the collection of systems provided in car for safety and comfort. FPGAs and now PSoC devices, can be used to realize these automotive systems \cite{fons2012fpga,velez2015reconfigurable}.

\textbf{- Image and Video Processing}: here, PSoCs processing capabilities are particularly valuable. Because they require both deterministic processing of large amount of pixel data, and software algorithms for extracting information from images \cite{dipert2012embedded}.

\textbf{- Medical}: An important issue in medical diagnosis is seeing inside the body. This task requires medical imaging equipment that requires sophisticated image processing algorithms to manage large data sets.  PSoCs offer capabilities that support both, high-speed parallel processing and software-based algorithms \cite{khan2012fpgas}.

\textbf{- High Performance Computing}: For fast processing of large datasets, which can typically be accelerated with dedicated hardware \cite{sundararajan2010high}. Recently, the growth of Deep Learning Systems has led an increase in the use of PSoCs in this field due to their massive parallel processing capacity and their high speed bus interfaces between PS and PL \cite{aimar2017nullhop,Qiu2016,Zhang2015,Zhang2016}.

In this paper a performance evaluation over a Xilinx PSoC memory transfers is presented and tested for a CNN accelerator application \cite{Aimar2016}. It consists on a user-level driver with several improvements, against a kernel-level driver. Both of them under a linux OS for embedded systems.

This paper is organized as follows. Section \ref{sec:psoc_platform} describes briefly Xilinx Zynq PSoC architecture and enumerates the interfaces between the PS (Processing System) and the PL (Programmable Logic) in the Zynq. Section \ref{sec:axidma_communication} explains the AXI DMA transfer flow at user-level and kernel-level drivers, while section \ref{sec:results} presents transfer timing results for each scenarios. Finally, section \ref{sec:conclusions} draws the conclusions.


\section{Xilinx PSoC Platform}\label{sec:psoc_platform}

Zynq chips from Xilinx are PSoCs architectures which contains an ARM-Cortex A family processor and a re-programmable logic (FPGA) in the same chip. A PSoC platform consists of a printed circuit board (PCB) that hosts a PSoC and several external chips to make the system to work properly under a Linux OS, typically. These external components are usually DDR memory, USB and Ethernet transceivers, SD card, JTAG for debugging and expansion connector with GPIOs.
These platforms represent a new co-design solution where the embedded OS (Linux) in the ARM cores executes software tasks (eg. data normalization, data recollection from sensors...) and a reconfigurable logic implements a design in order to accelerate a specific application. 

\begin{figure}[ht!]
\centering
\includegraphics[width=35mm]{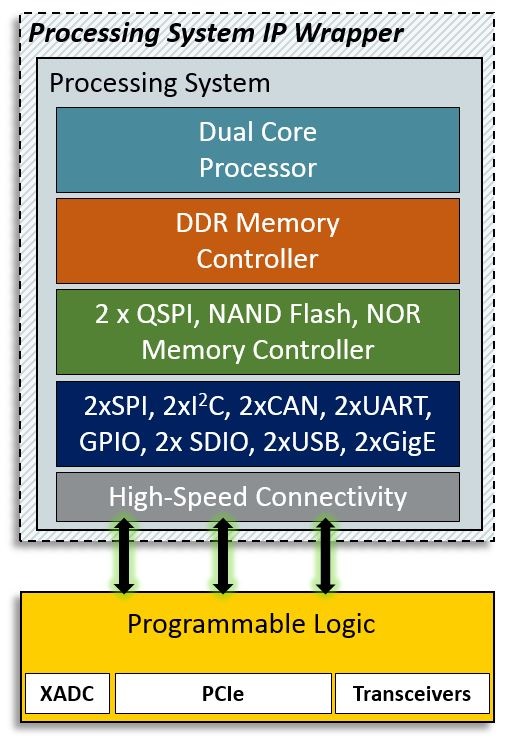}
\caption{Programmable logic communication with Processing System \label{fig1}}
\end{figure}

Interconnection between PL and ARM processor is done through PS. The PS is an ARM interface IP core, which acts as a logic connection between the ARM and the PL that assists to integrate custom and embedded IPs. This PS configures  different interfaces of the ARM core (I2C,SPI,..) and interfaces from PL, such as AXI, clock speed,... [Fig \ref{fig1}]. AXI stands for Advanced eXtensible Interface and the current version is AXI4, which is part of the ARM AMBA 3.0 open standard \cite{amba4axi4}. This AMBA standard was originally developed by ARM for microcontrollers but then it was extended for SoCs, including PSoCs, and it is an optimal interconnect technology between PS and PL. There are three different types of AXI4, each of which represents a different bus protocol, as summarized below: 

\textbf{- AXI4:} Oriented to memory-mapped links. It provides the highest performance. An address is supplied following by a data burst transfer of up 256 words (data word can be from 32 to 1024 bits) \cite{acasandrei2017open}.

\textbf{- AXI4-Lite:} A simplified link supporting only one data transfer per connection (no bursts). AXI4-Lite is also memory-mapped. In this case, an address and a single data word are transferred. This interface is commonly used to map control signals for devices \cite{acasandrei2017open}.

\textbf{- AXI4-Stream:} Oriented to high data flow applications with DMA support. It does not implement any handshake protocol. It allows unlimited burst transfers of unrestricted size. The protocol allows merging, packing and width conversion. It supports sparse, continuous, aligned and unaligned streams \cite{amba4axi4}.

\begin{figure}[ht!]
\centering
\includegraphics[width=80mm]{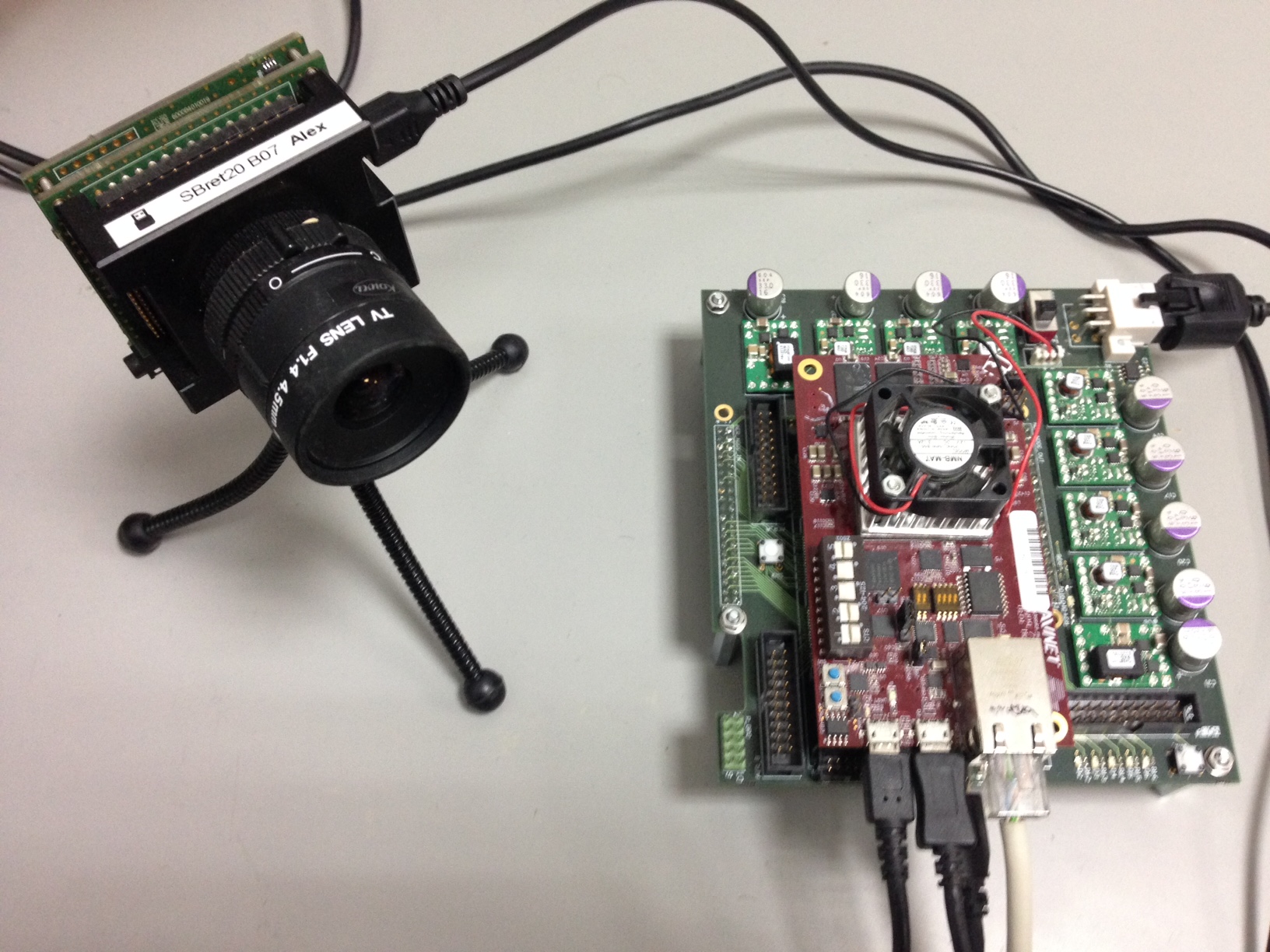}
\caption{Dock Soc platform and DAVIS  \label{figDockSoC}}
\end{figure}

In this work the Zynq-7100 MMP platform from Avnet has been used. This platform contains a PSoC with a Dual ARM\textsuperscript{\textregistered} Cortex\textsuperscript{\texttrademark}-A9 MPCore\textsuperscript{\texttrademark} operating at 666MHz with FPU engine, 1GB DDR3 memory, SD card support, USB and GigaEthernet. The PSoC includes a Kintex-7 FPGA with 444K logic cells in the same chip. Up to 132 GPIOs are available for external connectivity of the logic.
A baseboard, called DockSoC, designed for this MMP platform (manufactured by COBER) is able to manage all MMP needed power supplies (from 1V to 12V), the JTAG port over UART and several parallel interfaces to Neuromorphic chips over the CAVIAR and ROME parallel AER connectors are included \cite{serrano2009caviar}. The DockSoC can act as a daughter board for the AERNode\cite{aernode} platform to expand connectivity to other PSoC platforms and\/or to support the connectivity to other Neuromorphic systems. Figure \ref{figDockSoC} shows a picture of the used setup with the PSoC platform, the DockSoC baseboard and a USB neuromorphic retina, called DAVIS. The DAVIS \cite{davis} is a dynamic vision sensor that measures luminosity changes independently per pixel and send out events to signalize which pixel has detected such change in time over a configurable threshold. By collecting a fixed number of events from this sensor a histogram of those events can be used as a frame to be computed by the CNN accelerator running in the platform.

\section{AXI DMA communication}\label{sec:axidma_communication}


PL can be connected to the ARM processors by multiple interfaces as it was mentioned before. However, the fastest way is  using direct memory access (DMA) under the AXI Stream protocol, called AXI-DMA.
AXI-DMA consists of two different buses: Memory Mapped to Stream (MM2S) and Stream to Memory Mapped (S2MM). MM2S reads from DDR memory and transmits data to PL, while S2MM write data from PL to DDR memory. 
The DMA architecture presented in this paper contains two modules that have been created to adapt S2MM and MM2S interfaces data flow to and from the CNN accelerator implemented in the PL, called NullHop \cite{aimar2017nullhop}. 

NullHop is a hw accelerator designed for multi-layered CNNs execution for deep-learning classification applications. It resides in the PL and it needs to receive both the visual input (feature maps for a particular layer, or a portion of it) and the parameters (convolution kernels) from the PS, to calculate the results (output feature maps). It has been designed with 128 MAC blocks to work in a streamed way. Once the accelerator has received the parameters, the visual input is streamed in. After a couple of rows are received, the MACs start to operate and to produce an streamed output, which is sent back to the PS. To extract the maximum performance in our PSoC system, it is needed to properly coordinate the data flow in the application. When an OS is managing the PSoC, there are two different memory spaces: the virtual one, which where the user application works; and the physical one, which is managed by the DMA controller, and therefore, visible by the hardware implemented at the PL.

\begin{figure}[ht!]
\centering
\includegraphics[width=70mm]{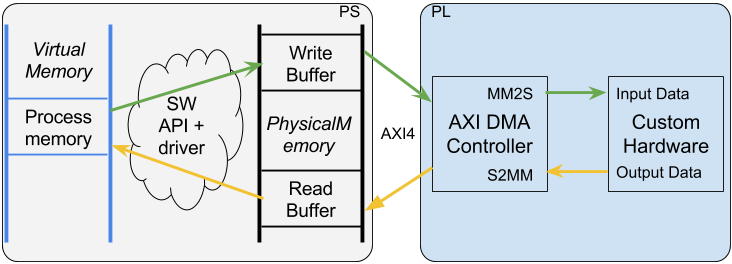}
\caption{Memory hierarchy in a PSoC with OS. User app works at virtual space, while DMA controller at PL works with physical one. The API and/or driver do the transfers to/from both spaces.
\label{virtual_physical_mem}}
\end{figure}

Figure \ref{virtual_physical_mem} shows the memory hierarchy from the user application to the CNN accelerator. Working with embedded Linux OS, there exist two ways to communicate with devices: (1) user-level: using  the function  mmap() to map a view of the device physical address space into our process virtual address space. This function is called by user application directly and the DMA transfers can be configured in a polling scheme, where the user application is frequently blocked, waiting for the transfer to be completed to process the data; or (2) kernel-level: a piece of software running at a higher privilege level of the OS, with interrupt support, in order to liberate the user application of blocking states until data is ready, allowing the execution of other needed tasks. Furthermore, the kernel-level ensures the integrity of the software avoiding the possible wrongly use of physical address spaces reserved to other processes running in the OS.

In this work a performance comparison between these two different communication schemes is presented. Furthermore, two different operating modes for the user-level driver have been introduced in the study: a completely polling-based solution, which would have the lowest latencies in between DMA transfers, and an scheduled solution, where DMA transfers are not continuously blocked.

\subsection{User-level} 
We have compared two read/write buffer implementations: single and double buffer. First one establishes only one channel for data transfers between virtual and physical memory. The double buffer implementation reserves two buffers in memory for virtual to physical transfers: while one is used for data ready to be sent to PL, the other one is used to prepare data for the next transmission. This second implementation allows reducing overhead latencies at OS level.
Apart from buffers implementation, two user-level driver operating modes have been implemented: \textit{Unique} and \textit{Blocks}.
\textit{Unique} mode sends all the data at once to the buffer, without any kind of partitioning. On the other hand, \textit{Blocks} mode divides data in smaller chunks of data for taking a better advantage of double buffering.  
Furthermore, two user-level versions have been compared: one completely based on polling, and a second one, closer to the kernel-level scenario explained in the following subsection, where a scheduler is managing the different DMA requests, to avoid dead-lock waits.


\subsection{Kernel-level}
In order to have the OS with a higher flexibility to attend other tasks for a realistic scenario, we have implemented a kernel-level driver that uses interrupts to manage the configuration of new DMA transfers when they are needed, allowing the PS to work in other tasks in the meantime. In this case, at user-level, the software specifies to the driver, at kernel-level, where all data are placed; then, the driver moves these data from virtual to physical space, and it configures the needed DMA transfers. In this case, we have used the AXI-DMA driver provided by Xilinx, which supports AXI-Stream DMA transfers with the needed length, or dividing them into small pieces and queuing them into consecutive transfers (Scatter-gated mode). To use the AXI-DMA Xilinx driver, a kernel-level API has been developed to adapt the driver to our needs. 

\section{Results}\label{sec:results}
\begin{figure}[ht!]
\centering
\includegraphics[width=85mm]{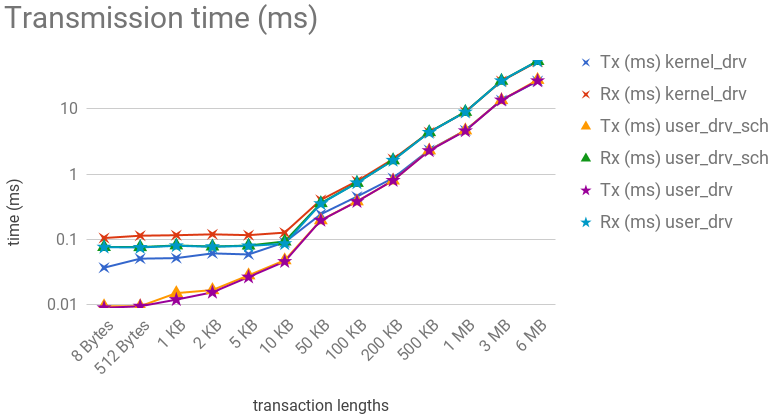}
\caption{Transfer times in ms for data blocks from 8B to 6MB comparing three drivers (user\_level, user\_level\_scheduled and kernel\_level).\label{TransmissionTime_ms}}
\end{figure}

\begin{figure}[ht!]
\centering
\includegraphics[width=85mm]{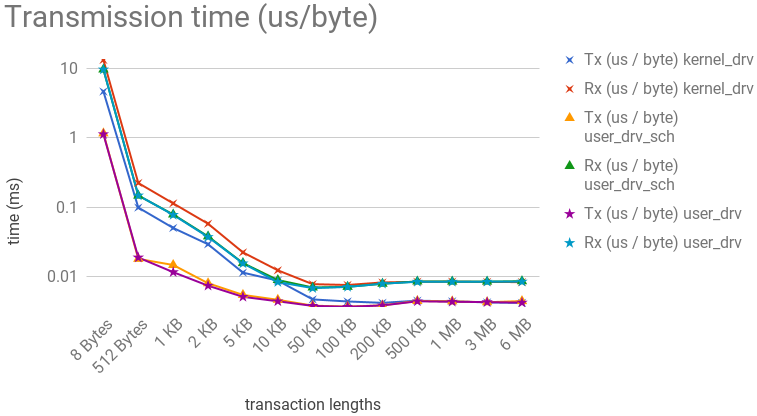}
\caption{Transfer times for 1byte (in us) for data blocks from 8bytes to 6MB comparing three drivers (user\_level, user\_level\_scheduled and kernel\_level).\label{TransmissionTime_us_byte}}
\end{figure}

We have tested the PSoC under two different scenarios: (1) with a hardware in a loop-back connection at PL that takes data from MM2S and stream it back to the S2MM interface of the DMA controller; and (2) a CNN execution using the NullHop accelerator at PL, executing the RoShamBo CNN \cite{aimar2017nullhop}.
Figures \ref{TransmissionTime_ms} and \ref{TransmissionTime_us_byte} shows the results for the first scenario. TX and RX transfer times evolution is presented for an incremental data size buffers from 8bytes to 6Mbytes considering the user-level driver with polling, the scheduled user-level and the interrupt-based kernel-level driver. 

For the loop-back streaming it could happen that TX and RX buffers would be full at the same time, so requests for reading RX buffers may occur at the same time that new TX request is produced. Since DDR memory cannot attend read and write operations at the same time, the bandwidth balance between RX and TX transfers is important in order to avoid blocking states of the system, eg. a longer enough TX transfers 
can fill up the RX hardware buffer and stops the TX transfer, blocking the system if 
RX and TX transfer are not properly managed. In these figures it can be seen that TX transfers have lightly higher priority than RX transfers, obtaining smaller latencies TX rather than RX transfers. Kernel-level driver approach, due to its bigger overhead at software execution because of the AXI-DMA Xilinx driver and the API, produces bigger latencies for smaller data lengths rather than user-level approach, but it increases the performance for bigger data lengths. User-level solution with polling and without scheduling gets lightly best results, but it could lacks on blocking the system while the transfers are done.

For the second scenario, we have set up the RoShamBo CNN execution in the MMP platform OS in the same way as described in \cite{aimar2017nullhop}, but we have modified the software to be using one of the three modes for controlling the memory transfers between virtual memory to physical memory and to manage the DMA transfer as described above. In this test, we have used the \textit{single-buffer} configuration and the \textit{Unique} mode. In table \ref{tab:table_nullhop} it can be seen the obtained timings for this case. The lowest latencies are obtained for the user-level mode with poling use. This is possible with this relative small CNN because transfer lengths are not longer enough to block the system. In \cite{aimar2017nullhop} bigger CNN were tested, such as VGG19, where this mode is not possible to be used and causes blocking the system. The second mode, without a kernel-level driver, but introducing a scheduler in the OS to avoid blocking the system, the latencies increases less than 2 ns per byte for TX and less that 150ns for RX. When the kernel driver is used, the latencies increases around 6 ns/byte for TX, but they decreases respect to the use of the scheduler, being less than 100ns slower than user-level. Regarding to the whole frame computation time, what requires the execution of 5 convolution layers in the NullHop, and therefore, sending and receiving DMA transfer for each layer; the latencies are bigger for the kernel-level driver, followed by the scheduler at user-level and then for the user-level. This behavior is correctly expected since transfer lengths for RoShamBo CNN are in the order of 100Kbytes, where kernel-level driver is still not obtaining its best results, as depicted in figures \ref{TransmissionTime_ms} and \ref{TransmissionTime_us_byte}.  

\begin{table}[]
\centering
\caption{CNN execution time for one frame and TX, RX average transfer times per byte}
\label{tab:table_nullhop}
\begin{tabular}{|l|l|l|l|}
\hline
\multicolumn{1}{|c|}{} & \multicolumn{3}{c|}{\cellcolor[HTML]{FFCC67}Unique mode, single-buffer} \\ \cline{2-4} 
\multicolumn{1}{|c|}{\multirow{-2}{*}{\textbf{NullHop RoShamBo}}} & \cellcolor[HTML]{FFCC67}TX (us/byte) & \cellcolor[HTML]{FFCC67}RX (us/byte) & \cellcolor[HTML]{FFCC67}Frame (ms) \\ \hline
\rowcolor[HTML]{ECF4FF} 
user-level polling & 0.0054 & 0.197 & 6.31 \\ \hline
\rowcolor[HTML]{9698ED} 
user-level drv scheduled & 0.0072 & 0.335 & 6.57 \\ \hline
\rowcolor[HTML]{00D2CB} 
kernel-level drv & 0.011 & 0.294 & 7.39 \\ \hline
\end{tabular}
\end{table}

\section{Conclusions}\label{sec:conclusions}
This paper presents and evaluates different implementations at software level of data movements between virtual memory space of an OS at user level, and physical memory space at kernel level for DMA transactions between the PS and the PL of a Xilinx Zynq PSoC for CNN executions. From the implementation at user-level privilege of the OS, using a polling solution, with less memory protection; to highest protection, using a kernel-level driver with interruptions; through an intermediate solution at user-level using an scheduler; this paper has evaluated two different scenarios, a real one under the execution of a CNN for playing RoShamBo with the NullHop CNN hardware accelerator, and a synthetic one for extracting the performance characteristics of the different implementations. 

User-level solutions give better latencies for data transfers bellow 1Mbyte, but they lacks on flexibility for multi-threading programs due to intensive use of polling. Their maximum supported transfer lengths are 8Mbytes (AXI4-Stream limit), but for big transfers the performance decreases due to long polling stages.

Kernel-level solution, tested for the worst possible case: single buffer scheme and unique data transfers, obtains similar latencies for bigger data transfer lengths. 
For the RoShamBo test, since transfer lengths are in the order of 100Kbytes, the user-level polling solution performs better due to have a smaller software overhead.

\section{Acknowledgment}
This work was partially supported by the NPP project funded by SAIT (2015-2018) and by the Spanish government grant (with support from the European Regional Development Fund) COFNET (TEC2016-77785-P). The work of R. Tapiador has been supported by a Formaci\'{o}n de Personal Investigador Scholarship from the University of Seville.

\bibliographystyle{IEEEtran}



\end{document}